\documentclass[twocolumn,showpacs,preprintnumbers,amsmath,amssymb,prl]{revtex4}

\usepackage{epsfig}

\newcommand{\T}{{\cal T}}

\begin{document}

\title{Duality relations for charge transfer statistics}
\author{A.~Komnik}
\affiliation{
Institut f\"ur Theoretische Physik, Universit\"at
Heidelberg, Philosophenweg 12, D-69120 Heidelberg, Germany}
\date{\today}

\begin{abstract}
By a detailed study of the mathematical structure of the cumulant generating function (CGF) of particle transfer in the non-interacting case we show that it satisfies duality relations connecting the 
CGFs of two different realizations of the same system. Employing these identities we derive an infinite number of relations between different cumulants, which amounts to a generalization of the fluctuation-dissipation theorem to non-equilibrium situations. We generalize the concept to multi-terminal set-ups and show that duality relations also exist for genuinely interacting systems by explicitly deriving one of them for the two-terminal anisotropic Kondo impurity model and discussing other important examples. The discovered identities are useful tools in the ongoing investigation of the analytical properties of the cumulant generating functions of charge transport.  
\end{abstract}

%\pacs{72.10.Fk, 72.25.Mk, 73.63.-b}

\maketitle

The exhaustive knowledge of the charge transfer properties of nanostructures is of paramount importance for their future applications. Not only such traditional quantities as non-linear current-voltage relations but also the current autocorrelations and noise spectra are required to ensure optimal operation of the devices in question. It turns out that current correlations of higher orders also carry important information and thus should also be considered \cite{reulet}. A very useful quantity encoding all these higher order correlation functions is the probability $P(Q)$ of transferring $Q$ particles through the device during the very long {\it waiting time} $\T$ \cite{Levitov1996,nazarov2009quantum}, also known as full counting statistics (FCS).  Its cumulant generating function (CGF) 
\begin{eqnarray}
 \ln \chi(\lambda) = \ln \sum_{Q=0}^N P(Q) e^{i \lambda Q} \, ,
\end{eqnarray}
is more convenient to work with. 
Here $\lambda$ is the counting field and $N$ is the maximally possible number of particles transferred through the system \footnote{It is also equal to the number of transferred particles for the perfectly transmitting device, of course.}. The irreducible moments (cumulants) $\langle Q^n \rangle$ of the order $n$  of this distribution function are obtained by a successive derivation with respect to counting field and by setting it to zero afterwards, 
\begin{eqnarray}
 \langle Q^n \rangle = (-i)^n \partial_\lambda \ln \chi(\lambda) |_{\lambda \to 0} \, .
\end{eqnarray} 
CGF is not a mere mathematical construction. Among other things it satisfies the fundamental Cohen-Gallavotti fluctuation theorem (CGFT) \cite{Evans1993,Gallavotti1995}, which enables to derive infinitely many relations between different cumulants of charge transport. The simplest of them is the celebrated Johnson-Nyquist relation between the first and the second cumulant. In this way CGFT offers a generalization of fluctuation-dissipation theorem to genuinely non-equilibrium situations. 

In view of the above it is very interesting to find out whether there are any additional relations. The goal of this Letter is to investigate this issue. By analyzing the transport of non-interacting particles first we derive a duality relation connecting the CGFs of two different realizations of the same system. We show that such identities also exist for genuinely interacting systems even in situations, when transport occurs by different kinds of quasiparticles. Moreover, we demonstrate how the duality relations can be used in order to access the transport properties of the systems in question and show avenues for further research.

%
%is the cumulant generating function (CGF) of the charge transfer statistics or full counting statistics (FCS) $\ln \chi(\lambda)$
%
%
%full counting statistics (FCS), which is mostly given in terms  

We start by considering a generic setup of an electronic device coupled to two metallic electrodes kept at chemical potentials $\pm e V/2$, where $V$ is the voltage applied to the device. First we consider non-interacting set-ups, in which the device is characterized by an energy dependent transmission coefficient $D(\omega)$ for single particles. It is well known that then the CGF is given by \cite{Levitov1996}
\begin{widetext}
\begin{eqnarray} 						\label{Eq1}
  \ln \chi[\lambda, v, D(\omega)] =
  G_0 \T \int d \omega \ln \left\{ 1 + D(\omega)
  \left[ n_L (1 - n_R) (e^{i \lambda} - 1) +  n_R (1 - n_L) (e^{-i \lambda} - 1) \right] \right\} \, ,
\end{eqnarray}
\end{widetext}
where $n_{L/R}= 1/[e^{(\omega \mp V/2)/T}+1]$ are the Fermi distribution functions for the electrons in the left/right (L/R) electrode, $T$ is the temperature in the system and $G_0 = e^2/h$ is the conductance quantum (here we neglect the spin degree of freedom and set $e=\hbar=k_B=1$ from now on).  The above equation can be rewritten as \cite{KomnikSaleur2011}:
\begin{eqnarray}                     \label{ch1}
&& \ln \chi[\lambda, v, D(\omega)] = G_0 {\cal T} \int d \omega \,
 \nonumber \\ &\times&
  \ln \left[ 1 + D(\omega) \, \frac{\cosh(v+i\lambda) - \cosh v}{\cosh(\omega/T) + \cosh v}\right]
% = \frac{T}{2\pi} \int d x \, \ln \left[\frac{\ch x + \ch w}{\ch x + \ch v} \right]
 \, ,
\end{eqnarray}
where $v=V/2T$. CGF in this form is particularly useful as it e.~g. helps to immediately recognize different symmetries. For instance, CGF is invariant under simultaneous change of sign in front of the voltage {\it and} of the counting field $\lambda$. The latter modification is essential as it changes the direction of particle counting, which is obviously necessary if one changes the direction of the applied voltage.

Far less obvious is the CGFT, which is the symmetry under the transformation 
\begin{eqnarray}					\label{CGtrafo}
\lambda \to -\lambda + i 2 v \, ,
\end{eqnarray} 
which also leaves the CGF unchanged. This property is easily traced back to the evenness of the hyperbolic cosine. One might now ask the question whether there are more transformations of this kind. Indeed, let us consider a similar change in the voltage
\begin{eqnarray}					\label{Noninttrafo}
 v \to - v - i \lambda \, . 
\end{eqnarray} 
Then we obtain 
\begin{eqnarray}
 &&\ln \chi[\lambda, - v - i \lambda, D(\omega)] = 
 - \ln \chi[\lambda,  v , 1] 
 \nonumber \\ \nonumber
 &+&
 G_0 {\cal T} \int d \omega \, \ln \left[ 
 1 + [1-D(\omega)] \frac{\cosh(v+i\lambda) - \cosh v}{\cosh(\omega/T) + \cosh v}\right] \, , 
\end{eqnarray}
where 
\begin{eqnarray}
 \ln \chi[\lambda,  v , 1] &=& G_0 {\cal T} \int d \omega \, \ln \left[ \frac{\cosh(\omega/T) +\cosh(v+i\lambda)}{\cosh(\omega/T) + \cosh v}\right] 
 \nonumber \\
 &=& G_0 {\cal T} \, T \, ( i \lambda 2 v - \lambda^2 ) 
\end{eqnarray}
is the CGF of a perfectly transmitting channel \cite{Levitov1996,TowardsFCS}. Therefore we obtain the following {\it CGF duality relation}:
\begin{eqnarray}						\label{Dual1}
  \ln \chi[\lambda, v, D(\omega)] &=&  \ln \chi[\lambda,  v , 1] 
  \nonumber \\
  &+& \ln \chi[\lambda, - v - i \lambda, 1 - D(\omega)] \, .
\end{eqnarray}
The physical content of this relation at least on the level of currents is quite lucid. 
While the lhs describes the particle transport with a certain transmission probability $D(\omega)$, the second term on the rhs is responsible for the backscattering of these particles, the respective probability being $1 - D(\omega)$. The sum of the transmitted and reflected particle streams should then be equal to the incoming stream, which is obviously described by the term $\ln \chi[\lambda,  v , 1]$. The second term on the rhs counts the backscattering events, that is why it is natural that the voltage sign is changed. This circumstance can be quantified by an explicit computation of the first cumulant: 
\begin{eqnarray}					\label{current}
\langle Q(v,D) \rangle &=& G_0 \T V + \langle Q(-v, 1-D) \rangle 
\nonumber \\
&=& G_0 \T V - \langle Q_{BS}(v, D) \rangle \, ,
\end{eqnarray}
where $\langle Q \rangle$ and $\langle Q_{BS} \rangle$ are the averages for the transmitted and backscattered charges.

On the contrary, the shift of the voltage by the counting field is non-trivial [see the last term in Eq.~\eqref{Dual1}] and starts playing an important role in the cumulants of higher order. For instance, for the second cumulant we obtain the following relation: 
\begin{eqnarray}					\label{noise}
 \langle Q^2(v,D) \rangle &=& 2 G_0 \T T +  \langle Q^2(-v,1-D) \rangle 
 \nonumber \\
 &+&
2 \partial_v  \langle Q(-v,1-D) \rangle \, ,
\end{eqnarray}
which relates the {\it non-linear} differential conductance in the backscattering channel, given by the very last term, to the second cumulants, or the noise. While the conventional fluctuation-dissipation relation connects the {\it linear} conductance to equilibrium fluctuation spectrum, our result holds for non-linear quantities. From the third cumulant onwards the perfect transmission term $\ln \chi [\lambda, v, 1]$ does not contribute any more and one finds the following universal interrelation between the cumulants of different higher orders: 
\begin{eqnarray}						\label{cumulantinterrelation}
 \langle Q^n (v, D) \rangle = \sum_{k=0}^{n-1} \left( \begin{array}{c}
 										n \\
										k
 									      \end{array} \right)
									      \partial_v^{(k)}  \langle Q^{n-k} (-v, 1- D) \rangle
\end{eqnarray}
We note in passing that the upper limit of the sum is essential. 
The identity \eqref{cumulantinterrelation} is one of our central results, from which a whole hierarchy of relations between different transport characteristics emerges. Thus its predictive power is comparable to that of the CGFT.

The existence of similar duality relations for {\it interacting} systems is not obvious. In the following we demonstrate that they are indeed possible. We exemplarily consider the transport through a single-channel two-terminal Kondo impurity, for which an explicit and quite compact formula for the CGF exists in the Toulouse limit \cite{TowardsFCS}. Here
\begin{widetext}
\begin{eqnarray}             \label{finalkondo}
&~&\ln\chi[\lambda, v, T_2,T_1] = G_0 {\cal T}  \int\limits_{0}^\infty d
\omega \, \ln \left\{1+T_2(\omega)
%\nonumber \right. \\  \nonumber &\times& \left.
\left[n_L(1-n_R)(e^{2i\lambda}-1)
+  n_R(1-n_L)(e^{-2i\lambda}-1)\right] \right. \\ \nonumber
&+& \left. T_1(\omega)\left[[n_F(1-n_R)+n_L(1-n_F)](e^{i\lambda}-1)
+
[n_F(1-n_L)+n_R(1-n_F)](e^{-i\lambda}-1)\right]\right\},
\end{eqnarray}
%\end{widetext}
%\begin{eqnarray}             \label{finalkondo}
%&~&\ln\chi[\lambda, v, T_1,T_2] = G_0 {\cal T}  \int\limits_{0}^\infty d
%\omega \, \ln \left\{1+T_2(\omega)
%\nonumber \right. \\  \nonumber &\times& \left.
%\left[n_L(1-n_R)(e^{2i\lambda}-1)
%+  n_R(1-n_L)(e^{-2i\lambda}-1)\right] \right. \\ \nonumber
%&+& \left. T_1(\omega)\left[[n_F(1-n_R)+n_L(1-n_F)](e^{i\lambda}-1)
%\right. \right. \\
%&+& \left. \left.
%[n_F(1-n_L)+n_R(1-n_F)](e^{-i\lambda}-1)\right]\right\},
%\end{eqnarray}
where $T_{1,2}$ are effective transmission coefficients for single particles and particle pairs, which depend on the parameters of the model, and $n_F = 1/(e^{\omega/T} + 1)$ is the plain Fermi distribution function. First it is desirable to recast this result into the form similar to Eq.~\eqref{ch1}:
\begin{eqnarray}    				\nonumber 
 \ln\chi[\lambda, v, T_2,T_1] = G_0 {\cal T}  \int\limits_{0}^\infty d
\omega \ln \left\{ 1+T_2(\omega) \, \frac{\cosh(v+i 2 \lambda) - \cosh v}{\cosh(\omega/T) + \cosh v}
+ 2 T_1(\omega) \, \frac{\cosh(v+i\lambda) + \cosh(i \lambda) - \cosh v - 1}{\cosh(\omega/T) + \cosh v}
\right\} \, .
\end{eqnarray}
%\end{widetext}
%\begin{eqnarray}
% &~&\ln\chi[\lambda, v, T_1,T_2] = G_0 {\cal T}  \int\limits_{0}^\infty d
%\omega \\ \nonumber 
%&\times& \ln \left\{ 1+T_2(\omega) \, \frac{\cosh(v+i 2 \lambda) - \cosh v}{\cosh(\omega/T) + \cosh v}
%\right. \\ \nonumber 
%&+& \left.
%2 T_1(\omega) \, \frac{\cosh(v+i\lambda) + \cosh(i \lambda) - \cosh v - 1}{\cosh(\omega/T) + \cosh v}
%\right\} \, .
%\end{eqnarray}
Despite the presence of two different channels one immediately verifies that the CG symmetry \eqref{CGtrafo} is perfectly satisfied. In the next step we perform a transformation 
\begin{eqnarray}					\label{Kondotrafo}
 v \to - v - i 2 \lambda
\end{eqnarray} 
inspired by Eq.~\eqref{Noninttrafo}. It results in
%\begin{widetext}
\begin{eqnarray}
 &~&\ln\chi[\lambda, -v - i 2 \lambda, T_2,T_1] = - \ln \chi[\lambda,  v , 1]  + G_0 {\cal T} \nonumber  \\ \nonumber 
&\times&  \int\limits_{0}^\infty d
\omega \ln \left\{ 1+[1-T_2(\omega) - 2 T_1(\omega)] \, \frac{\cosh(v+i 2 \lambda) - \cosh v}{\cosh(\omega/T) + \cosh v}
+
2 T_1(\omega) \, \frac{\cosh(v+i\lambda) + \cosh(i \lambda) - \cosh v - 1}{\cosh(\omega/T) + \cosh v}
\right\} \, .
\end{eqnarray}
%\begin{eqnarray}					\label{Kondotrafo}
% v \to - v - i 2 \lambda \, . 
%\end{eqnarray} 
\end{widetext}
%\begin{eqnarray}
% &~&\ln\chi[\lambda, -v - i 2 \lambda, T_1,T_2] = - \ln \chi[\lambda,  v , 1] + G_0 {\cal T}  \int\limits_{0}^\infty d
%\omega \nonumber  \\ \nonumber 
%&\times& \ln \left\{ 1+[1-T_2(\omega) - 2 T_1(\omega)] \, \frac{\cosh(v+i 2 \lambda) - \cosh v}{\cosh(\omega/T) + \cosh v}
%\right. \\ \nonumber 
%&+& \left.
%2 T_1(\omega) \, \frac{\cosh(v+i\lambda) + \cosh(i \lambda) - \cosh v - 1}{\cosh(\omega/T) + \cosh v}
%\right\} \, .
%\end{eqnarray}
and leads to the following duality relation: 
\begin{eqnarray}						\label{DualKondo}
 && \ln\chi[\lambda, v, T_2,T_1] =  \ln \chi[2 \lambda,  v , 1] 
  \nonumber \\
  &+&\ln\chi[\lambda, -v - i 2 \lambda, 1-T_2-2 T_1,T_1]\, .
\end{eqnarray}
Needless to say, also here the relations \eqref{current}-\eqref{cumulantinterrelation} still apply.

A completely different transport problem for interacting systems is that of tunnelling between two Luttinger liquids (e.~g. between two counterpropagating edge states in the fractional quantum Hall device) via a resonant level \cite{Kane1992}. Contrary to the Kondo set-up, in which the interaction is localized on the impurity here the electrons are interacting in the bulk of the electrodes. For interacting strength $g=1/2$ (or filling fraction $\nu=1/2$) there is a simple analytical solution for the CGF in a closed form, which is given by an equation similar to Eq.~\eqref{finalkondo} \footnote{Here one has to set  $T_1=0$.} \cite{Saleur2001,KT,KondoDotFCS}. Interestingly, the corresponding duality relation \eqref{DualKondo} connects the seemingly different results reported in Refs.~\cite{KT,KondoDotFCS} and was indirectly derived by other means in \cite{TowardsFCS}. We would also like to mention that a different kind of duality has been pointed out in \cite{Saleur2001}. It connects the CGFs for forward- and backscattered particles, but at strictly zero temperature. 

Yet another interacting system in which a duality relation exists is a normal metal-superconductor tunnelling contact. Its CGF was computed in \cite{Muz1994}. It is easiest to see for the case of energies below the gap, when Andreev reflections dominate the transport. Then the transformation \eqref{Kondotrafo} generates the corresponding relation. The doubled counting field is important here as it reflects the effective charge of current carrying quasiparticles.

It is not difficult to generalize the above concept to set-ups with $N$ different terminals. For the CGF there is a very useful result \cite{nazarov2009quantum}
\begin{eqnarray} 					\label{LLL}
 \ln \chi(\lambda, \mu, s) = G_0 \T \int d \omega \, \mbox{Tr} \ln \left[
 1 + {\bf f} (s^\dag \widetilde{s} - 1) 
 \right]	
 \nonumber \\	
 = G_0 \T \int d \omega \,  \ln \mbox{det} \left[
 1 + {\bf f} (s^\dag \widetilde{s} - 1)
 \right]
  \, ,
\end{eqnarray}
where $\lambda = (\lambda_1, \dots, \lambda_N)$ and $\mu = (\mu_1, \dots, \mu_N)$ are the sets of counting fields and chemical potentials in the respective terminals, ${\bf f}=\mbox{diag}(n_1, \dots, n_N)$ is a diagonal matrix of Fermi distribution functions, $s$ is the (energy-dependent) scattering matrix of the system, and $\widetilde{s}_{i j} = s_{i j} e^{i (\lambda_i - \lambda_j)}$. Let us now assume that there is a realization of the same system with a different scattering matrix $s_0$. Then we can rewrite the above determinant as
\begin{eqnarray}
\mbox{det} \left[
 1 + {\bf f} (s^\dag \widetilde{s} - 1)
 \right] = \mbox{det} \left[
 1 + {\bf f} (s_0^\dag \widetilde{s_0} - 1)
 \right] \nonumber \\
 \times
 \mbox{det} \left[
 1 + {\bf g}
 \left({\bf B} -1 \right)
 \right] \, ,
\end{eqnarray}
where ${\bf g} = [1+{\bf f} (s_0^\dag \widetilde{s}_0 - 1)]^{-1} {\bf f} s_0^\dag \widetilde{s}_0$ and ${\bf B} = (s_0^\dag \widetilde{s}_0)^{-1} s^\dag \widetilde{s}$. The above relation leads to a factorization of the charge transfer distribution function and a {\it separation} of the CGF into two parts: 
\begin{eqnarray}					\label{Nchannels}
\ln \chi(\lambda, {\bf f}, s^\dag \widetilde{s}) =  \ln \chi(\lambda, {\bf f}, s_0^\dag \widetilde{s}_0) + 
\ln \chi(\lambda, {\bf g}, {\bf B}) \, .
\end{eqnarray}
Such kind of CGF separation is obviously always possible. A duality is more restrictive and emerges  from \eqref{Nchannels} when two conditions are met: (i) ${\bf g}$ can be obtained from ${\bf f}$ by a $\lambda$-dependent shift of chemical potentials \footnote{${\bf g}$ has to be non-singular, but not necessarily diagonal because one always can choose a diagonal ${\bf h} = {\bf U} {\bf g} {\bf U}^{-1}$, so that det$\left[
 1 + {\bf g}
 \left({\bf B} -1 \right)
 \right]=\mbox{det} \left[
 1 + {\bf h}
 \left( {\bf U} {\bf B} {\bf U}^{-1} - 1 \right)
 \right]$.} ; (ii) for ${\bf B}$
there is a factorization ${\bf B}=s_1^\dag \widetilde{s_1}$ with some meaningful $s_1$. 

For instance, for $N=2$ and $s_0$ describing perfect transmission one finds for $\lambda_1=-\lambda_2=\lambda/2$, $\mu_1=-\mu_2=V/2$, that  ${\bf g}(v)={\bf f}(-v-i\lambda)$ just as proposed in Eq.~\eqref{Noninttrafo}. ${\bf B}$ is equivalent to $s_1^\dag \widetilde{s_1}$, where $s_1$ is obtained from $s$ by interchanging
the reflection and transmission amplitudes. In total one then recovers the result \eqref{Dual1}. The multi-terminal duality is not restricted to systems with two terminals only. It was shown in \cite{TowardsFCS}, that the result \eqref{finalkondo}
can be obtained by recasting the transport problem into a four-terminal geometry and subsequently using the formula \eqref{LLL}.

It is interesting to discuss the implications of duality relations to the analytical properties of the CGF. 
At zero temperature and in the case of structureless transmission $D(\omega)=D$ of non-interacting particles the CGF is given by
\begin{eqnarray}    					\label{basicstart}
 \ln \chi[\lambda,V,D] = G_0 {\cal T} V \, \ln \left[ 1 + D ( e^{i \lambda}-1) \right] \, .
\end{eqnarray}
For $D<1/2$ this function can be expanded with respect to the small parameter $D e^{i \lambda}/(1-D)$ and is thus a periodic function of $\lambda$ with period $2\pi$. For $1/2< D < 1$ the singularity of the log can be reached in \eqref{basicstart} leading to a discontinuity of the imaginary part at $\lambda$ of multiples of $\pi$. A proper analytical continuation is done via another expansion after rewriting the CGF as
\begin{eqnarray}					\label{Dual0}
&& \ln \chi[\lambda,V,D] = i  \lambda G_0 {\cal T} V \nonumber \\ &+& G_0 {\cal T} V \, \left[\ln D  + \ln \left( 1 + \frac{1-D}{D} \, e^{- i \lambda} \right) \right] \, ,
\end{eqnarray}
and expanding in terms of $(1-D) e^{-i \lambda}/D$. Obviously, the latter function is not periodic but has a linear in $\lambda$ contribution. So the duality relation \eqref{Dual0} accomplishes the analytical continuation of the CGF into the other transmission regime.

We have shown that the cumulant generating function of charge transfer satisfies duality relations connecting the full counting statistics of two different realizations of the same system. These duality relations turn out to be useful in deriving additional identities for different cumulants of particle transfer statistics thereby generalizing and extending the fluctuation-dissipation relations to non-equilibrium systems. These relations turn out not only to hold for non-interacting particles but also for transport through Kondo impurities, superconducting contacts and tunnelling between interacting one-dimensional systems. Importantly, the concept we put forward can be generalized also to multi-terminal set-ups with such interesting implications as factorization of the distribution functions and separability of the cumulant generating functions. 
It would be very interesting to investigate the issue of existence of such duality relations at finite temperature in systems for which no analytical expressions in closed form exist. Furthermore, an analysis of spin-resolved dualities also appears to be a valid avenue for further research. 

The author is supported by the Heisenberg Programme of the Deutsche Forschungsgemeinschaft (Germany) under Grant No. KO 2235/5-1.
\bibliography{CGFduality}

\end{document}